\documentclass[manuscript]{emulateapj}
\usepackage{amsmath}
\usepackage{amssymb}
\usepackage{natbib}
\usepackage{hyperref}
\usepackage{breakurl}
\usepackage{xcolor}
\usepackage{enumitem}
\setlist{leftmargin=5.5mm}
\usepackage{lineno}

\newcommand{\mail}{lilirayhk@phys.ncku.edu.tw}
\newcommand{\flux}{\,erg\,cm$^{-2}$\,s$^{-1}$}
\newcommand{\lum}{\,erg\,s$^{-1}$}

\newcommand{\cm}{\,cm$^{-2}$}
\newcommand{\nh}{$N_\mathrm{H}$}
\newcommand{\src}{4FGL~J0336.0+7502}

\shorttitle{\src\ as a black widow}
\shortauthors{Li et al.}

\begin{document}
\title{Revealing a new black widow binary \src}
\author{
Kwan-Lok Li\altaffilmark{1},
Y. X. Jane Yap\altaffilmark{2},
Chung Yue Hui\altaffilmark{3}, and
Albert K. H. Kong\altaffilmark{2}
}

\altaffiltext{1}{Department of Physics, National Cheng Kung University, 70101 Tainan, Taiwan; \href{mailto:\mail}{\mail} (KLL)}
\altaffiltext{2}{Institute of Astronomy, National Tsing Hua University, Hsinchu 30013, Taiwan}
\altaffiltext{3}{Department of Astronomy and Space Science, Chungnam National University, Daejeon, Republic of Korea}

\begin{abstract}
We report on a discovery of a promising candidate as a black widow millisecond pulsar binary, \src, which shows many pulsar-like properties in the 4FGL-DR2 catalog. Within the 95\% error region of the LAT source, we identified an optical counterpart with a clear periodicity at $P_{\rm orb}=3.718178(9)$~hours using the \textit{Bohyunsan 1.8-m Telescope}, \textit{Lulin One-meter Telescope}, \textit{Canada–France–Hawaii Telescope}, and \textit{Gemini-North}. At the optical position, an X-ray source was marginally detected in the \textit{Swift}/XRT archival data, and the detection was confirmed by our \textit{Chandra}/ACIS DDT observation. The spectrum of the X-ray source can be described by a power-law model of $\Gamma_x=1.6\pm0.7$ and $F_{\rm 0.3-7keV}=3.5^{+1.2}_{-1.0}\times10^{-14}$\flux. The X-ray photon index and the low X-ray-to-$\gamma$-ray flux ratio (i.e., $<1$\%) are both consistent with that of many known black widow pulsars. There is also a hint of an X-ray orbital modulation in the \textit{Chandra} data, although the significance is very low ($1.3\sigma$). If the pulsar identity and the X-ray modulation are confirmed, it would be the fifth black widow millisecond pulsar binary that showed an orbitally-modulated emission in X-rays. 
\end{abstract}

\section{Introduction}
The \textit{Fermi Large Area Telescope} (LAT) has been observing the MeV/GeV $\gamma$-ray sky since June 2008. 
With the ten years of data taken between 2008 and 2018, 5064 sources are detected in the \textit{Fermi} LAT 10-Year Point Source Catalog (4FGL-DR2; \citealt{2020ApJS..247...33A,2020arXiv200511208B}). While a major portion of the cataloged sources are known systems (e.g., active galaxies, pulsars, etc.), about one-fourth of them are unidentified at other wavelengths. Other than active galaxies, many of these unidentified $\gamma$-ray sources are believed to be pulsar systems. 

There have been multi-wavelength searching campaigns conducted for new candidates of $\gamma$-ray pulsars from the list of the unidentified \textit{Fermi}-LAT sources (e.g., \citealt{2015ApJ...809...68H,2020MNRAS.497.5364B}). Machine learning techniques were also applied on the classification for pulsars based on only the $\gamma$-ray properties recently (e.g., \citealt{2016ApJ...820....8S,2020MNRAS.492.5377L,2020MNRAS.495.1093H}). 
These efforts have led to at least a dozen candidates for further radio/$\gamma$-ray pulsation searches. Many of these candidates could be associated with two special pulsar classes, black widow (BW) and redback (RB), which are millisecond pulsars in compact 
binaries (the orbital periods are often less than a day). 
Besides the compact orbits, the two classes are characterised by the very low-mass companions (i.e., 0.1--0.4 $M_\odot$ for RBs and $<0.1M_\odot$ for BWs; \citealt{2013IAUS..291..127R,2013ApJ...775...27C}) ablated by the strong radiations that originate from the primary pulsars. 
The radiation would also heat up the tidally-locked companion one-sided, and this so-called pulsar heating effect can result in orbital modulations in the optical bands (see, e.g., \citealt{2016ApJ...828....7R,2019A&A...621L...9Y} for the details), although exceptions exist (e.g., 3FGL~J0212.1+5320 that does not show any observable pulsar heating effect; \citealt{2016ApJ...833..143L}).
Some recently discovered BW/RB candidates include 3FGL~J0954.8$-$3948 \citep{2018ApJ...863..194L}, 4FGL~J2333.1$-$5527 \citep{2020ApJ...892...21S}, 4FGL~J0935.3+0901 \citep{2020MNRAS.493.4845W}, 4FGL~J0407.7$-$5702 \citep{2020ApJ...904...49M}, and 4FGL~J0940.3$-$7610 (\citealt{2021ApJ...909..185S}; see the Table 3 of \citealt{2019Galax...7...93H} and the references therein for more candidates). 

In this paper, we present a multi-wavelength study for \src, which is a new BW MSP candidate identified by our unidentified \textit{Fermi}-LAT sources observing campaign. The study includes (i) the \textit{Fermi}-LAT $\gamma$-ray properties of \src\ (\S\ref{sec:fermi}); (ii) optical photometric observations taken by the \textit{Bohyunsan 1.8-m Telescope}, \textit{Lulin One-meter Telescope} (LOT), \textit{Canada–France–Hawaii Telescope} (CFHT), and \textit{Gemini-North} (\S\ref{sec:opt}); (iii) \textit{Swift}/XRT and \textit{Chandra}/ACIS-S X-ray observations (\S\ref{sec:xray}); and (iv) a discussion for \src\ based on its multi-wavelength properties (\S\ref{sec:discuss}).

\begin{table*}
\scriptsize
\centering 
\caption{The 4FGL-DR2 \texttt{LogParabola} parameters of \src\ and ten arbitrarily chosen black widow millisecond pulsars.}
\begin{tabular}{lccccl}
\toprule
Name		&	$\Gamma_\gamma$	&	$\beta_\gamma$ \\
\hline
\src\ & $1.78\pm0.10$ & $0.39\pm0.07$ \\
PSR~J0023+0923&	$1.89\pm0.11	$& $0.46\pm	0.09$\\
PSR~J0610$-$2100&	$1.99\pm0.09$&	$0.51\pm	0.06$\\
PSR~J0952$-$0607&	$2.02\pm0.18$&	$0.68\pm	0.11$\\
PSR~J1124$-$3653&	$1.89\pm0.06$&	$0.34\pm	0.04$\\
PSR~J1301+0833&	$1.93\pm0.08$&	$0.41\pm	0.06$\\
PSR~J1544+4937&	$2.07\pm0.10$&	$0.20\pm	0.05$\\
PSR~J1731$-$1847&	$2.33\pm0.24$&	$0.71\pm	0.25$\\
PSR~J1805+0615&	$2.01\pm0.14$&	$0.50\pm	0.12$\\
PSR~J2047+1053&	$1.70\pm0.15$&	$0.39\pm	0.08$\\
PSR~J2214+3000&	$1.68\pm0.04$&	$0.49\pm	0.03$\\
\hline
\label{tab:4fgl_bw}
\end{tabular}
\end{table*}

\section{Gamma-ray properties}
\label{sec:fermi}
\src\ is a bright $\gamma$-ray source (detection significance of $31.1\sigma$) located at a Galactic latitude of $b=15.5\arcdeg$ \citep{2020arXiv200511208B}. It was first discovered in $\gamma$-rays by \textit{Fermi}-LAT as 1FGL~J0334.2+7501 in the 1FGL catalog \citep{2010ApJS..188..405A}, and was subsequently cataloged in 2FGL, 3FGL, and 4FGL(-DR2) \citep{2012ApJS..199...31N,2015ApJS..218...23A,2020ApJS..247...33A,2020arXiv200511208B}. In 4FGL-DR2 (using data taken from August 2008 to August 2018), \src\ was classified as a steady source on a yearly time-scale with a variability index\footnote{A source with a variability index greater than 18.48 has less than 1\% chance to be stable} of 9.8. The average $\gamma$-ray flux in 100~MeV--100~GeV is $F_\gamma=(7.4\pm0.5)\times10^{-12}$\flux\ that is among the top 30\% of all 4FGL-DR2 sources. According to 4FGL-DR2, the $\gamma$-ray spectrum in the LAT energy domain is significantly curved with $>8\sigma$, and can be described by a \texttt{LogParabola} model, 

\begin{equation}\nonumber
\begin{split}
\dfrac{dN}{dE} \propto \Big(\frac{E}{E_b}\Big)^{-(\Gamma_\gamma+\beta_\gamma\log(E/E_b))},
\end{split} 
\end{equation}
where $\Gamma_\gamma$ is the photon index, $\beta_\gamma$ defines the degree of curvature, and $E_b$ is a fixed scale parameter. The best-fit spectral parameters are $\Gamma_\gamma=1.78\pm0.10$
and $\beta_\gamma=0.39\pm0.07$. In Table \ref{tab:4fgl_bw}, we have listed the $\Gamma_\gamma$ and $\beta_\gamma$ parameters of ten arbitrarily chosen black widow millisecond pulsars in 4FGL-DR2 for comparison.

\begin{figure*}
\centering
\includegraphics[width=1.0\textwidth]{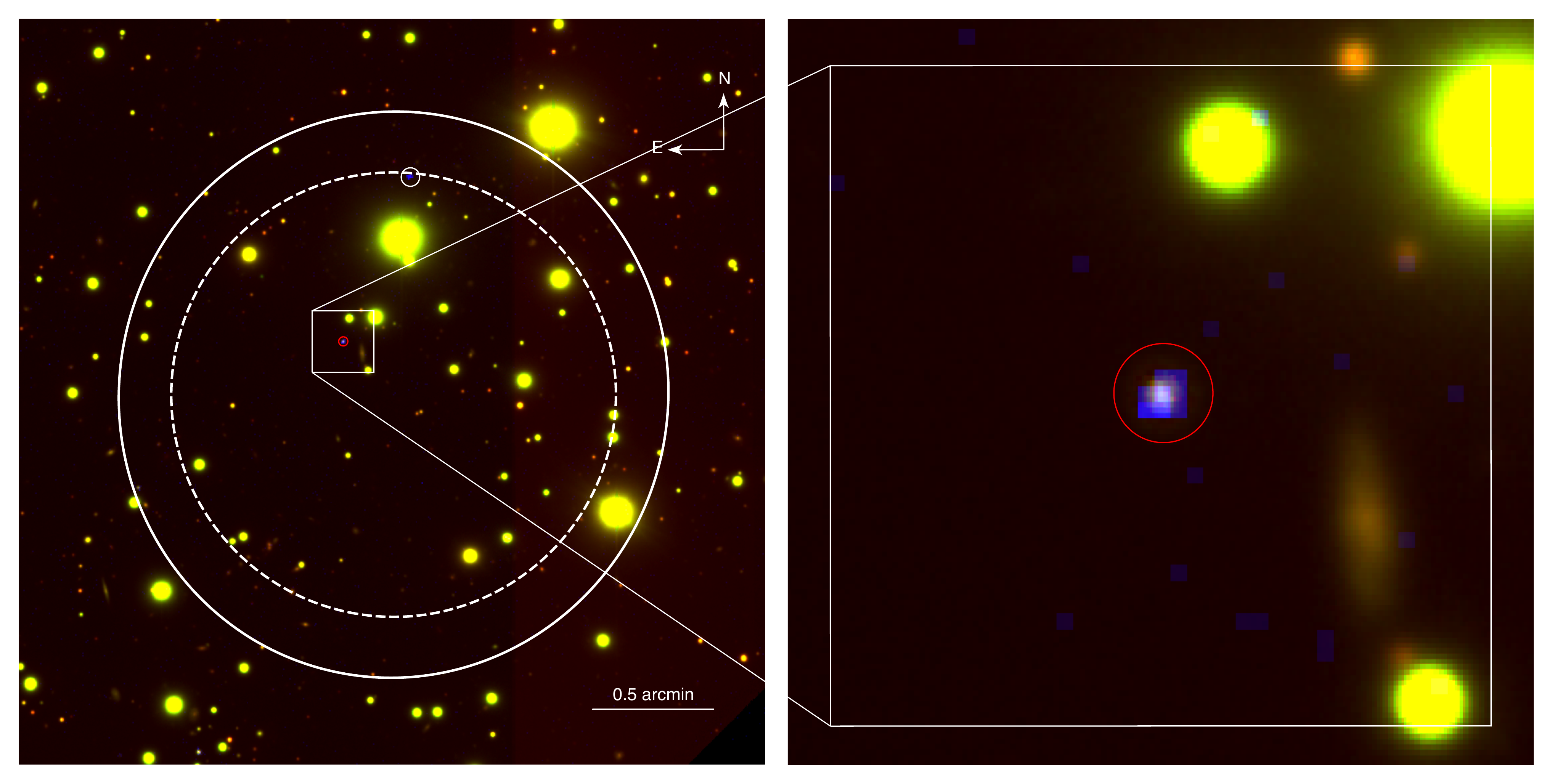}
\caption{False-color image (red: \textit{Gemini-North} \textit{i'} band; green: \textit{Gemini-North} \textit{r'} band; and blue: \textit{Chandra}/ACIS 0.3--7~keV) of the field of \src. The large white ellipse shows the \textit{Fermi}-LAT 95\% error ellipse of \src\ (about 1\farcm5 in radius) and the dashed one is the 68\% version (about 1\farcm2 in radius). The small red circle indicates the proposed optical/X-ray counterpart to the $\gamma$-ray source and the small white circle in the north direction shows another X-ray source detected by \textit{Chandra} in the field.}
\label{fig:finder}
\end{figure*}

The $\gamma$-ray properties of \src, including the off-plane Galactic latitude, the low long-term variability, and the significant spectral curvature, are exactly the characteristics that are commonly seen in $\gamma$-ray pulsars (see, e.g., \citealt{2015ApJ...809...68H}). In fact, \src\ has been suggested by \cite{2016ApJ...820....8S} as a good MSP candidate using machine learning techniques. This motivated us to investigate the system further in multi-wavelength. With the high quality of the LAT data, the 95\% positional accuracy of \src\ is around 1\farcm5, which is fine enough to allow a feasible search for the counterpart in optical and X-rays (see Figure \ref{fig:finder}).

\section{Optical photometric light curves}
\label{sec:opt}
\subsection{\textit{Bohyunsan 1.8-m Telescope}}
On 2019 March 31, we observed \src\ with the 1.8-m telescope at \textit{Bohyunsan Optical Astronomy Observatory} (BOAO) as part of our multi-wavelength observing campaign for unidentified \textit{Fermi}-LAT sources. The idea is to blind search for compact binaries in the field and carry out further multi-wavelength observations for them, if any. 

\src\ was observed 19 times in the \textit{R} band with the BOAO 4k CCD camera. The exposure time is 200 seconds for each image. \texttt{IRAF} was used for standard data reduction processes, including bias, dark, and flat calibrations. Aperture photometry was employed to extract light curves, and differential technique was applied to eliminate the variations due to the changing weather condition. We then calibrated the obtained magnitudes using the Fifth USNO CCD Astrograph Catalog (UCAC5; \citealt{2017AJ....153..166Z}).

Within the \textit{Fermi}-LAT 95\% error circle, we spotted a variable star that was undetected in the first 9 frames, and then became observable with the magnitude raising from $R\approx23$~mag to $\approx21$~mag in just 40 minutes. We stacked the first 9 images, and the variable was marginally detected with $R\approx24$~mag (Figure \ref{fig:opt_lc}). The variable source is also cataloged in the \textit{Gaia} DR3 with $\alpha\mathrm{(J2000)}=03^\mathrm{h}36^\mathrm{m}10\fs1811(1)$, $\delta\mathrm{(J2000)}=+75\arcdeg 03\arcmin 17\farcs268(1)$ (54.0424214(5), 75.0547967(3); \citealt{2020arXiv201201533G}), at which an X-ray source was marginally detected in archival \textit{Swift}/XRT observations (see \S\ref{sec:swift}). 

\begin{figure*}
\centering
\includegraphics[width=0.49\textwidth]{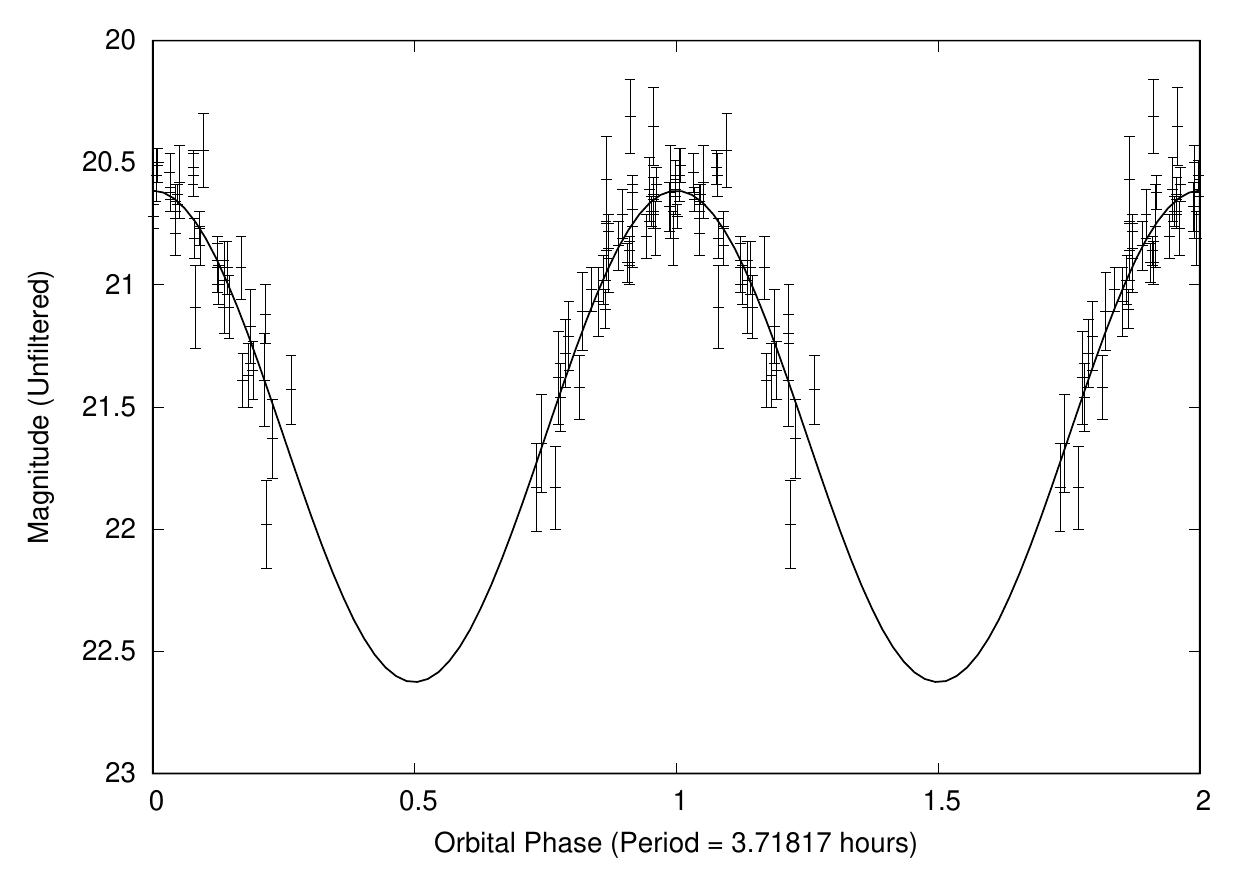}
\includegraphics[width=0.5\textwidth]{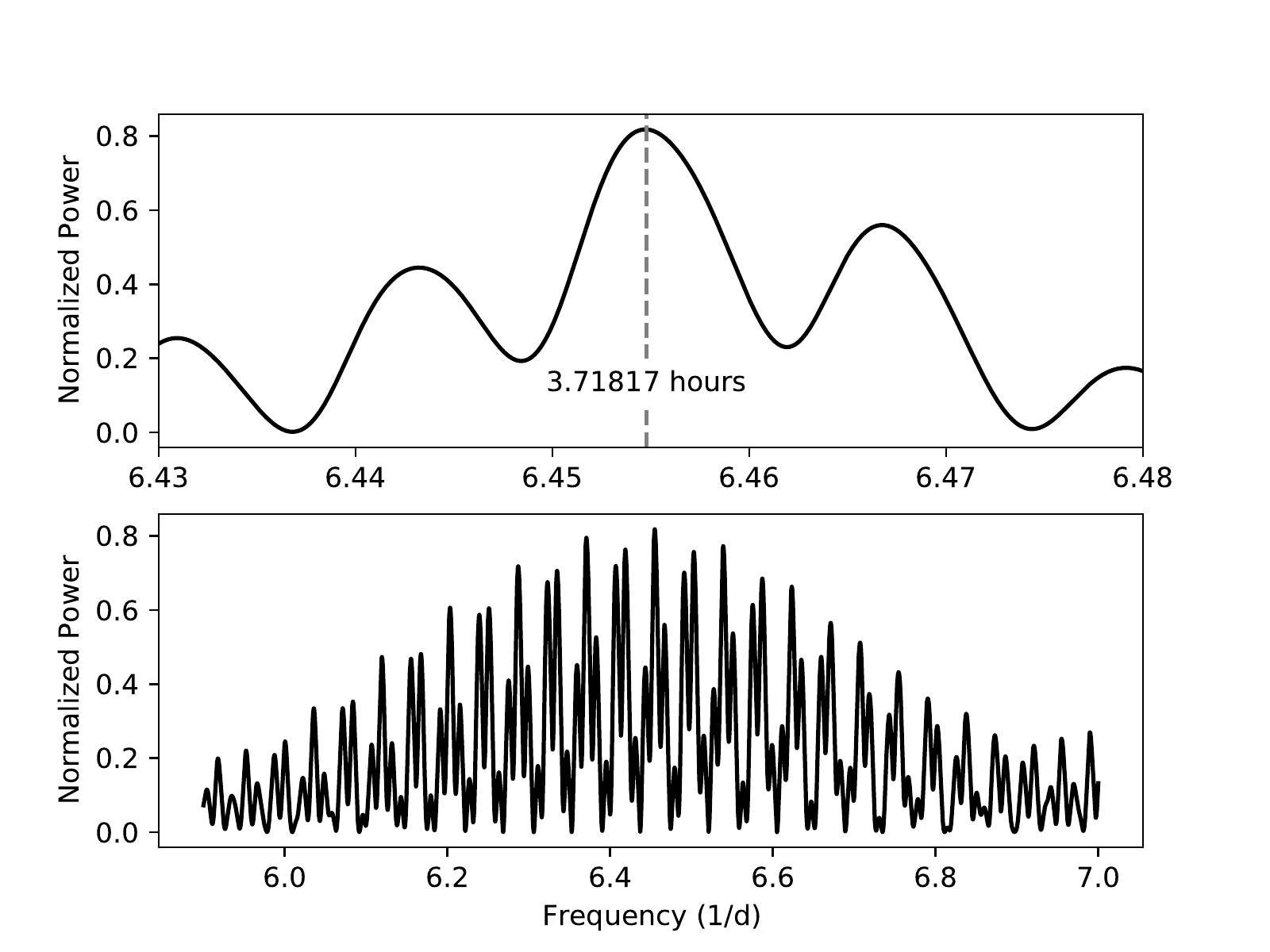}
\caption{Left: The phased optical LOT light curve of \src. Phase zero corresponds to the peak of the phased optical light curve (i.e., the superior conjunction of the companion, if it is a pulsar system). The solid line shows the best-fit sinusoidal function of the data. Two identical cycles are shown for clarity. Right: The Lomb-Scargle periodogram of the LOT data (bottom) and a close-up of the global peak (top).}
\label{fig:lulin}
\end{figure*} 

\subsection{\textit{Lulin One-meter Telescope}}

We followed-up the variable source using LOT at \textit{Lulin Observatory}. The observing dates are 2019 October 17/18, 2019 November 10, and 2020 January 8/9. As the source is faint for a 1-m class telescope, the data were taken unfiltered with 600 seconds per frame. 174 frames were taken in total in the 5 days. Standard data reduction and analysis processes were carried out using \texttt{IRAF} as described in the previous section. Flux calibration was done using the Pan-STARRS catalog (PS1 DR2; \citealt{2020ApJS..251....7F}), although the observations are unfiltered. 

The target's brightness was varying in the LOT light curve as we have seen in the BOAO data. About half of the LOT observations therefore result in non-detections when the source was in the faint phase. Despite the incompleteness, the LOT light curve shows a clear periodic modulation on a time-scale of about 4 hours. We computed a Lomb-Scargle periodogram for the LOT data (the non-detections were all rejected) and a strong signal was found at 3.7182 hours, albeit with aliases due to the non-detection gaps (Figure \ref{fig:lulin}). This periodicity is likely the orbital period of the system. We fitted the LOT light curve with a sinusoidal function to fine tune the orbital period ($P_{\rm orb}$). 
The tuned timing parameters are $P_{\rm orb,lot}=3.71817$~hours and $T_{\rm 0,lot}=2458858.016$ (BJD; converted by the method presented in \citealt{2010PASP..122..935E}). The latter is the epoch of phase zero, which is defined as the time when the phased light curve peaks. 
This should refer to the superior conjunction of the companion (i.e., observer-pulsar-companion along the line of sight), if it is a pulsar binary. In this convention, the ascending node of the pulsar and the inferior conjunction of the companion (i.e., observer-companion-pulsar) are at phases 0.25 and 0.5, respectively, for a circular orbit. 
The phased LOT light curve and the best-fit sinusoid are shown in Figure \ref{fig:lulin}. 

We also combined the LOT data with the BOAO, CFHT, and \textit{Gemini-North} $R/r^{\prime}$-band light curves (which will be presented in the following sections) to make a further improvement on the solution. For the $R/r^{\prime}$-band light curves, only data points brighter than 22~mag were selected as the fainter parts are not consistent with a sinusoid. The finalised timing solution of \src\ is $P_{\rm orb}=3.718178(9)$~hours and $T_{\rm 0}=2458858.0150(2)$, of which the uncertainties are 10 times smaller than that of the LOT solution. Figure \ref{fig:opt_lc} shows the LOT, BOAO, CFHT, and \textit{Gemini-North} light curves folded with the improved orbital solution. 

\subsection{\textit{Canada–France–Hawaii Telescope}}
On 2017 January 25 and February 2, we used CFHT/MegaCam on Mauna Kea to observe \src\ under our snapshot observing program. Two $g^\prime$-band and two $r^\prime$-band images were taken on the first night, and four $g^\prime$-band and two $r^\prime$-band images were taken in the second run. Except for the first two $g^\prime$-band observations on the second night that were exposed 30 seconds each, every image has an exposure time of 860 seconds. Pre-processed CFHT observations by the standard \texttt{Elixir} pipeline \citep{2004PASP..116..449M} were directly used. Flux calibration and data analysis methods were the same as the ones for the LOT data.

\begin{figure*}
\centering
\includegraphics[width=0.7\textwidth]{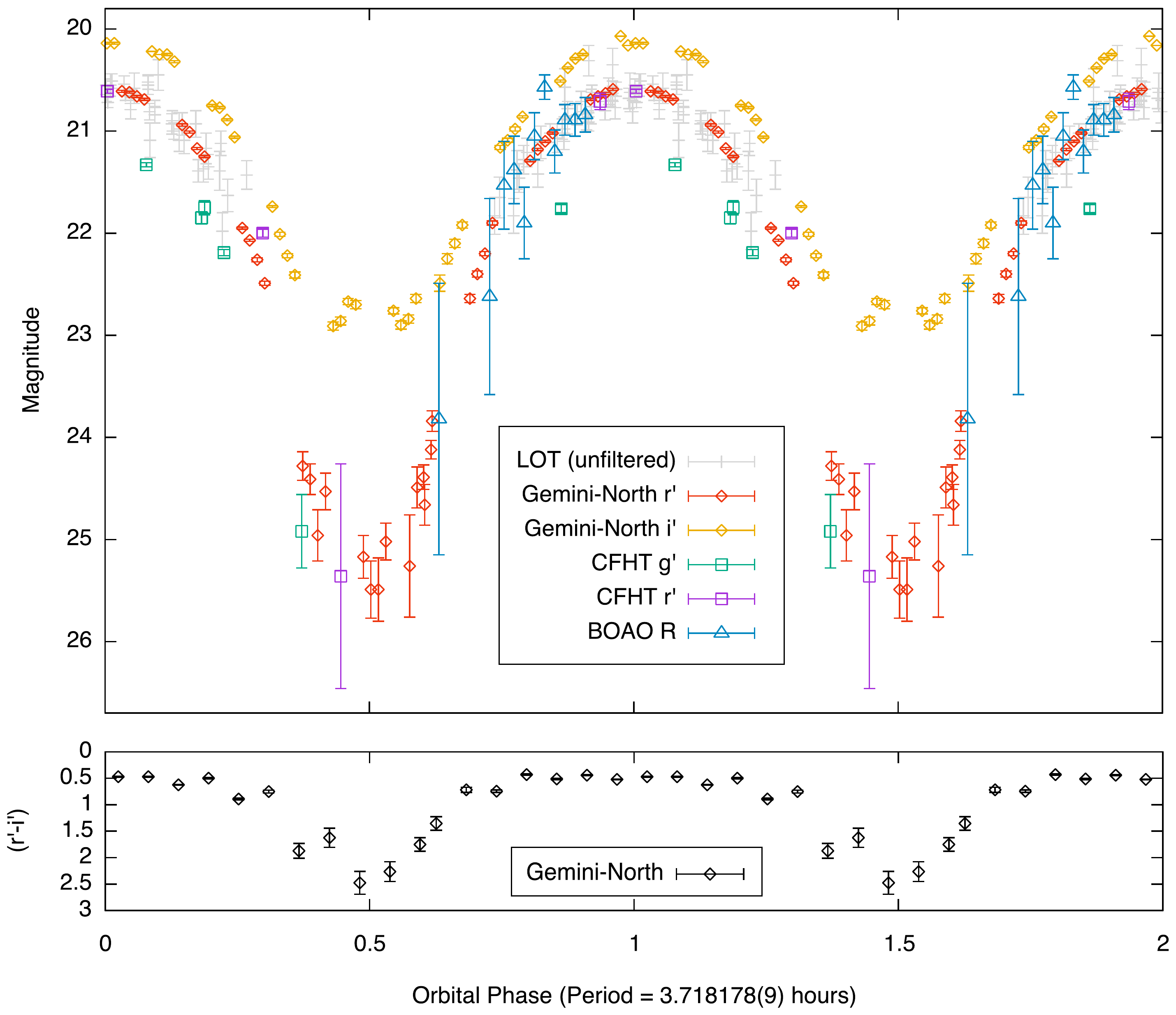}
\caption{The phased optical light curves (top) and color index curve (bottom) of \src. Phase zero corresponds to the peak of the phased optical light curve (i.e., the superior conjunction of the companion, if it is a pulsar system). Two identical cycles are shown for clarity.}
\label{fig:opt_lc}
\end{figure*}

The optical counterpart was clearly detected in all ten CFHT observations and the faintest measurements are around 25 mag in both bands. Folded with the global timing solution, both $g^\prime$- and $r^\prime$-band light curves show clear modulations that are in phase with the BOAO and LOT light curves (the green and the purple data points in Figure \ref{fig:opt_lc}). This shows that the binary has been stable over the last 3 years. 

\subsection{\textit{The Frederick C. Gillett Gemini Telescope}}
We also proposed a 4-hour series of \textit{Gemini-North} photometric observations (PI: Kwan-Lok Li) to study the complete modulation profile of \src. The observations were all taken using GMOS-N on 2019 November 28. Two filters, $r^\prime$ and $i^\prime$, were used with 150 seconds for each exposure. We set four observations as a turn to alternate between the filters. We intended to combine the observations taken in the same turn when the source became too faint to be detected in a single image, but it turns out that the source was clearly seen in all the individual images. Standard data reduction was done using \texttt{DRAGONS} \citep{2019ASPC..523..321L}. 
Differential photometry was used to extract the light curves, and the magnitudes were calibrated against the PS1 DR2 catalog. 

The light curves were also folded with the orbital solution and the orbital modulations are clearly shown in both bands (Figure \ref{fig:opt_lc}). The GMOS-$r^\prime$ light curve is well consistent with the ones observed with CFHT-$r^\prime$ and BOAO-$R$. While the peak-to-peak amplitude is around 5~mag in the $r^\prime$ band, the $i^\prime$-band amplitude is only 3~mag, indicating a significant orbital color variation (i.e., the source becomes redder as it is fainter), which is likely caused by pulsar heating. The modulation profiles are mostly smooth, but the $i^\prime$-band emission was significantly varying in the order of 0.1~mag around phase 0.5. In the $r^\prime$ band, the variation can also be seen around phase 0.5, although it is less significant. 
The variations, taken as a whole, look like mini-flares on time-scales of 300--900 seconds.

\section{X-ray observations}
\label{sec:xray}
\subsection{\textit{Neil Gehrels Swift}/XRT}
\label{sec:swift}
In the \textit{Swift}'s public data archive, there are four archival \textit{X-Ray Telescope} (XRT) observations taken for \src\ between 2012 April 14 and 18. The total effective exposure of the XRT data is 9~ksec. In the 0.3--10~keV XRT image, a weak X-ray source was seen at the position of the proposed optical counterpart. 9 photons were detected within a 15\arcsec\ radius circular source region centred at the optical source. Using a 120\arcsec\ source-free circular background region, we computed the average background counts within the source region to be 0.9 counts. This gives a net count rate of $\sim9\times10^{-4}$~counts/s with a detection significance of $>9\,\sigma$. Despite the low photon statistics, we adopted the \textit{online XRT product generator}\footnote{\url{https://www.swift.ac.uk/user_objects/index.php}} to extract the X-ray spectra (binned to at least one count per bin) and used XSPEC (v12.11.1) to estimate the spectra parameters using an absorbed power-law model. The column density (\nh) was fixed at the Galactic value of $1.47\times10^{21}$\cm\ \citep{2016A&A...594A.116H} and the \textit{Cash statistic} \citep{1979ApJ...228..939C} was employed as the fitting statistic. The best-fit photon index is $\Gamma_x=0.9^{+3.3}_{-2.5}$ with a corrected X-ray flux of $F_{\rm 0.3-10keV}=5.1^{+4.9}_{-3.9}\times10^{-14}$\flux\ (all the uncertainties are in 90\% confidence interval). 

\begin{figure}
\centering
\includegraphics[width=0.47\textwidth]{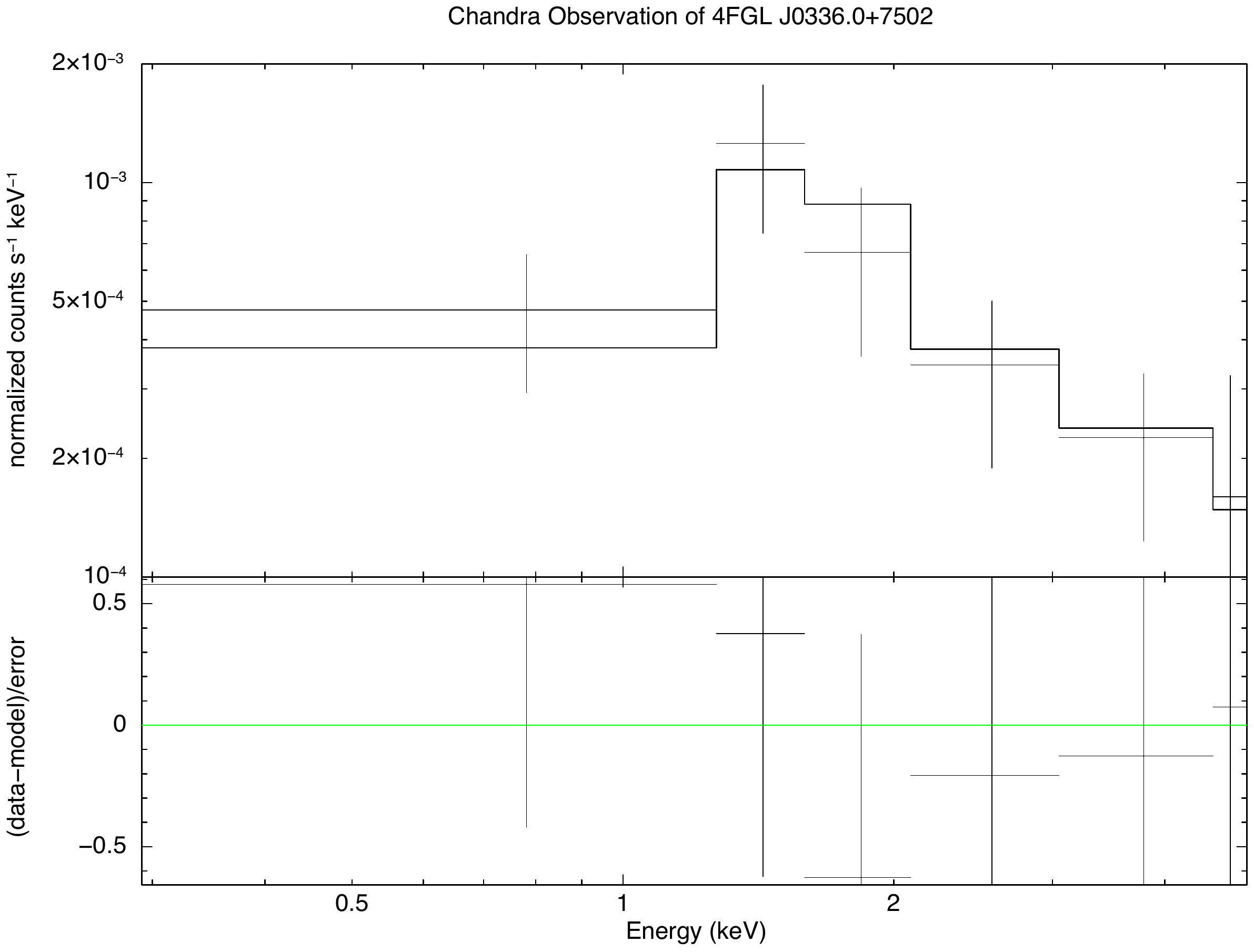}
\caption{The \textit{Chandra}/ACIS X-ray spectrum with its best-fit absorbed power-law model (\nh =$1.47\times10^{21}$\cm, $\Gamma_x=1.6\pm0.7$, and $F_{\rm 0.3-7keV}=3.5^{+1.2}_{-1.0}\times10^{-14}$\flux). While we used at least one count per bin for the model fitting, the displayed spectrum was binned to 5 counts per bin (except the last bin) for better visualization.}
\label{fig:chandra_spec}
\end{figure} 

\begin{figure}
\centering
\includegraphics[width=0.49\textwidth]{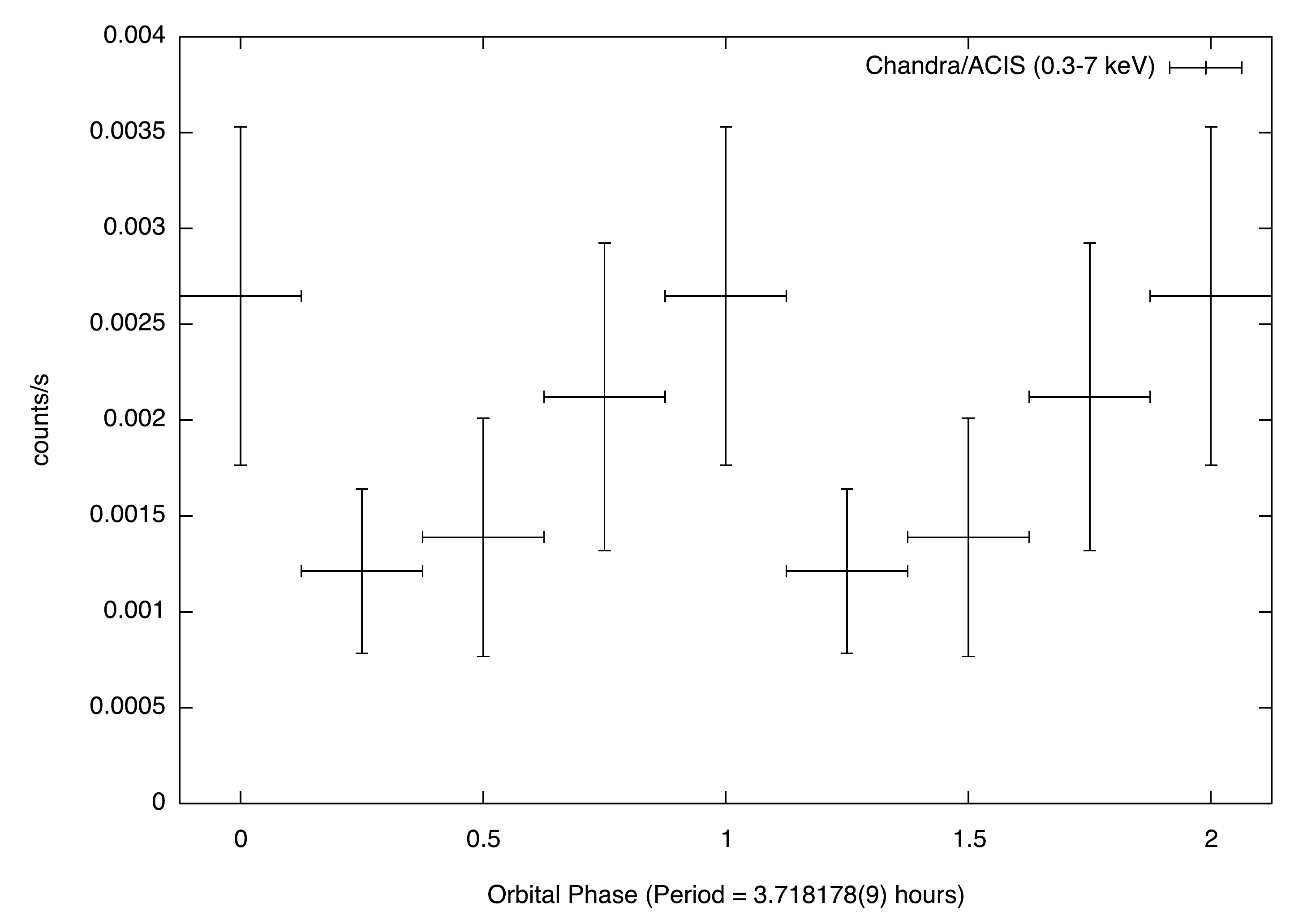}
\caption{The phased \textit{Chandra}/ACIS X-ray light curve of \src. Phase zero corresponds to the peak of the phased optical light curve (i.e., the superior conjunction of the companion, if it is a pulsar system). Two identical cycles are shown for clarity.}
\label{fig:chandra_lc}
\end{figure} 

\subsection{\textit{Chandra}/ACIS}
We requested and obtained a 15-ksec \textit{Chandra} Director's Discretionary Time (DDT) observation for \src\ to confirm the XRT detection, study the full-orbit X-ray modulation, and constrain the spectral parameters better. The observation was taken on 2020 January 25 with the Advanced CCD Imaging Spectrometer S-array (ACIS-S; \citealt{2003SPIE.4851...28G}) operated in the full frame mode. 

\texttt{CIAO} (v4.12) with \texttt{CALDB} (v4.9.0) was used to reduce and analyse the data. After reprocessing the data using the \texttt{CIAO} script \texttt{chandra\_repro}, the proposed X-ray counterpart to \src\ was clearly detected in the 0.3--7~keV band. We employed \texttt{specextract} and \texttt{dmextract} with a 1\farcs5 radius source region and a source-free annulus background region centered at the target (inner/outer radii: 5\arcsec/10\arcsec) to extract the X-ray spectrum and light curve of the source, respectively. We also applied a barycentric correction on the X-ray light curve using \texttt{axbary}. 

A total of 29 photons were extracted in the 0.3--7~keV band, and about 1\% of them are from the background. This gives an average net count rate of $\sim2\times10^{-3}$~counts/s. Given the insufficient number of source counts for a detailed spectral analysis, we simply binned the \textit{Chandra} spectrum to at least one count per bin using \texttt{grppha} and fitted an absorbed power-law model (\nh\ was fixed to $1.47\times10^{21}$\cm) to it with \textit{Cash statistic} using \texttt{XSPEC} (Figure \ref{fig:chandra_spec}). The best-fit parameters are $\Gamma_x=1.6\pm0.7$ and $F_{\rm0.3-7keV}=3.5^{+1.2}_{-1.0}\times10^{-14}$\flux\ (90\% confidence interval), which are consistent with the values obtained from the \textit{Swift}/XRT observations.

We folded the X-ray light curve using the optical timing solution. Since the number of counts is low, the phased light curve has only 4 data bins per cycle and each bin has about 7 counts on average. While the signal-to-noise ratios of the light curve are low, it shows a possible X-ray modulation that seemingly peaks at phase zero. However, the light curve is also consistent with a flat count rate of $\sim2\times10^{-3}$~counts/s (Figure \ref{fig:chandra_lc}). Assuming a flat light curve, the $\chi^2$ value is 4.7 with 3 degrees of freedom, equivalent to a chance probability of 19\% or $1.3\sigma$. Except the epoch folding, the light curve quality does not allow any further investigations, such as X-ray hardness studies. 

Besides the proposed X-ray counterpart, there is another brighter X-ray source (total number of counts is 58) detected within the 95\% 4FGL-DR2 error ellipse and at the edge of the 68\% error region (Figure\ref{fig:finder}). Its coordinates are $\alpha\mathrm{(J2000)}=03^\mathrm{h}36^\mathrm{m}04\fs6$, $\delta\mathrm{(J2000)}=+75\arcdeg 04\arcmin 10\farcs6$ (90\% uncertainty circle\footnote{\url{https://cxc.harvard.edu/cal/ASPECT/celmon/}}: 0\farcs8), at which no optical counterpart can be identified in the \textit{Gemini-North} images (the limiting magnitude $\approx$ 26~mag in the $r'$ band). Assuming an absorbed power-law model (a fixed value of \nh\ $=1.47\times10^{21}$\cm\ was assumed), the best-fit spectral parameters are $\Gamma_x=1.9\pm0.4$ and $F_{\rm 0.3-7keV}=7.0^{+1.9}_{-1.5}\times10^{-14}$\flux. No strong variability can be seen in a 1000-sec binned X-ray light curve. As there is already compelling evidence to show that the fainter X-ray source (presented in last paragraph) is most likely the counterpart to \src, we suggest that this brighter X-ray source is an unrelated system and no further discussion will be given in this paper.

\section{Discussion and Conclusion}
\label{sec:discuss}
We have identified a likely optical and X-ray counterpart to \src, which is an X-ray binary with an orbital period of 3.718178(9) hours. Together with the $\gamma$-ray properties (see \S\ref{sec:fermi}) and the low X-ray-to-$\gamma$-ray flux ratio (i.e., $\lesssim 1$\%; see Table \ref{tab:bw}) that are all consistent with a BW/RB MSP binary, we propose that \src\ is a new system of the class. 

While the optical counterpart to \src\ has been cataloged in \textit{Gaia} DR3, the source is faint for \textit{Gaia} (i.e., $G=20.6$~mag), and hence, no distance information can be extracted through the parallax measurement. We estimated the distance to \src\ by assuming $L_x\lesssim10^{32}$\lum\ that is a typical limit for BW/RB pulsars in a rotation-powered state \citep{2018ApJ...864...23L}. The inferred distance is $d\lesssim5$~kpc, which results in a very faint absolute magnitude of $M_{r^\prime}\gtrsim11.5$~mag (with no pulsar heating). Therefore, \src\ would highly likely be a BW (instead of RB) MSP binary. Indeed, the optical orbital modulation of \src\ is very much close to that of other BW systems (e.g., PSR~J1311$-$3430; \citealt{2012ApJ...760L..36R}). 

\begin{table*}
\scriptsize
\centering 
\caption{X/$\gamma$-ray properties of \src\ and the BW MSPs with X-ray orbital modulation known in the field.}
\begin{tabular}{lccccl}
\toprule
Name		&	Variability Index\footnote{ A source with a variability index greater than 18.48 has less than 1\% chance to be stable.}		&	Spectral Curvature\footnote{ The fit improvement over a simple power-law (PL) model when a power law with exponential cut-off (PLEC) is assumed.}	&	Photon Index 		&	$L_x/L_\gamma$\footnote{ The energy ranges of the X-ray and $\gamma$-ray luminosities are 0.1--100~GeV and 0.3--8~keV, respectively.}	&	References \\
		&	 ($\gamma$-ray)	&	($\gamma$-ray; $\sigma$)		&	(X-ray)	&	(\%)			&		 \\
\hline
\src\footnote{ The X-ray modulation is just marginally seen.}			&	9.8		&	8.6		&	1.6	&	0.5 & (this work)\\
PSR J1124$-$3653\footnotemark[4]	&	8.1		&	11.0		&	1.3	&	0.5 & \cite{2014ApJ...783...69G}\\
PSR J1653$-$0158\footnotemark[4] & 9.2 & 14.5 & 1.6 & 0.5 & \cite{2014ApJ...794L..22K,2020ApJ...902L..46N}\\
PSR J1959+2048	&	11.5		&	10.8		&	2.0	&	0.4 & \cite{2012ApJ...760...92H}\\
PSR J2256$-$1024	&	18.7		&	7.4		&	1.8	&	0.5 & \cite{2014ApJ...783...69G}\\
\hline
\label{tab:bw}
\end{tabular}
\end{table*}

Besides the optical brightness, the $(r^\prime-i^\prime)$ color index changes over the orbit. In general, the color is much redder around the inferior conjunction, and this is a signature of pulsar heating. In addition to the periodic variations, there were a few mini-flares detected around the inferior conjunction. It is unclear whether the flares actually concentrate at the inferior conjunction as they might distribute evenly and just became prominent when the optical brightness of the companion was lowest. It is worth noting that flaring activities have been recently shown common in BW/RB systems in both the rotation/accretion-powered states (e.g., \citealt{2017ApJ...850..100A,2018ApJ...858L..12P,2018MNRAS.477.1120K,2019A&A...621L...9Y,2020ApJ...895...89L}), although the physical origin is still under debate. 

An insignificant orbital modulation ($1.3\sigma$) is also seen in X-rays. If confirmed, it would be the fifth BW MSP system that exhibits an orbital modulation in X-rays (Table \ref{tab:bw}). The X-ray orbital modulation could be caused by the Doppler boosting along the intrabinary shock flow \citep{2014ApJ...785..131T,2014ApJ...797..111L}, and thus the profile strongly depends on the wind momentum ratio between the stellar wind of the companion and the pulsar wind, $\beta$. For BWs, $\beta<1$ (i.e., a much weaker stellar wind) is expected \citep{2016ApJ...828....7R}, and the X-ray emission should peak around the inferior conjunction (i.e., phase 0.5 for the definition used in Figure \ref{fig:chandra_lc}). However, the X-ray orbital modulation profile of \src\ is different than expected--the X-ray peak is at the superior conjunction--and this might imply a stronger stellar wind (i.e., $\beta>1$). According to \cite{2019Galax...7...93H}, there are three BW MSPs (one of them is the original BW system, PSR~J1959+2048) that possibly have orbitally-modulated X-ray emission \citep{2012ApJ...760...92H,2014ApJ...783...69G}. Including PSR~J1653$-$0158 that was identified as a candidate pulsar system by \cite{2014ApJ...794L..22K} and recently confirmed as a BW by the GPU-accelerated Einstein@Home \citep{2020ApJ...902L..46N}, four BW systems are known to have (possible) X-ray orbital modulations (Table \ref{tab:bw}). Surprisingly, PSR~J1959+2048 is the only one with a phased X-ray light curve consistent with the case of $\beta<1$. All three others, as we have seen in \src, are more consistent with $\beta>1$. This might subvert our general impression on the BWs' companions that their winds are not strong. 
Alternatively, the X-ray modulation could be caused by occultation, if the shock region is small and very close to the companion instead of forming a bow shock which wraps around the star. In this case, $\beta<1$ is still possible. However, shape dips as the light curve minima would be expected because of the very tiny size of the companion star in a black widow system. This does not match very well with the X-ray observations that show broad minima in the light curves (e.g., Figure \ref{fig:chandra_lc}). 
Future deep X-ray observations of these systems could confirm/deny these ``abnormal'' X-ray modulations and/or provide helpful hints to solve the problem. 

\begin{acknowledgements}
KLL is supported by the Ministry of Science and Technology of the Republic of China (Taiwan) through grant 109-2636-M-006-017, and he is a Yushan (Young) Scholar of the Ministry of Education of the Republic of China (Taiwan).
YXJY and AKHK are supported by the Ministry of Science and Technology of the Republic of China (Taiwan) through grant 109-2628-M-007-005-RSP.
C.Y.H. is supported by the National Research Foundation of Korea through grants 2016R1A5A1013277 and 2019R1F1A1062071.

This publication has made use of data collected at Lulin Observatory, partly supported by MoST grant 108-2112-M-008-001.

Based on observations obtained with MegaPrime/MegaCam, a joint project of CFHT and CEA/DAPNIA, at the Canada-France-Hawaii Telescope (CFHT) which is operated by the National Research Council (NRC) of Canada, the Institut National des Science de l'Univers of the Centre National de la Recherche Scientifique (CNRS) of France, and the University of Hawaii. The observations at the Canada-France-Hawaii Telescope were performed with care and respect from the summit of Maunakea which is a significant cultural and historic site.

Based on observations obtained at the international Gemini Observatory, a program of NSF’s NOIRLab, which is managed by the Association of Universities for Research in Astronomy (AURA) under a cooperative agreement with the National Science Foundation. on behalf of the Gemini Observatory partnership: the National Science Foundation (United States), National Research Council (Canada), Agencia Nacional de Investigaci\'{o}n y Desarrollo (Chile), Ministerio de Ciencia, Tecnolog\'{i}a e Innovaci\'{o}n (Argentina), Minist\'{e}rio da Ci\^{e}ncia, Tecnologia, Inova\c{c}\~{o}es e Comunica\c{c}\~{o}es (Brazil), and Korea Astronomy and Space Science Institute (Republic of Korea).

This work made use of data supplied by the UK Swift Science Data Centre at the University of Leicester.

The scientific results reported in this article are based on observations made by the Chandra X-ray Observatory. This research has made use of software provided by the Chandra X-ray Center (CXC) in the application packages CIAO, ChIPS, and Sherpa.

\end{acknowledgements}
\textit{Facilities}: \facility{BOAO:1.8m, LO:1m, Gemini:Gillett, CFHT, CXO, and Swift}

\bibliography{j0336}
\end{document}